# Brain Stroke Classification Using Wavelet Transform and MLP Neural Networks on DWI MRI Images

Mana mohammadi, Amirhesam Jafari Rad, Ashkan Behrouzi

Email: {mana.mohammadi, a.h.jafarirad, ashkan.behrouzi}@ut.ac.ir

*Electrical and Computer Engineering*
*University of Tehran*
Tehran, Iran

*Abstract*— This paper presents a lightweight framework for classifying brain stroke types from Diffusion-Weighted Imaging (DWI) MRI scans, employing a Multi-Layer Perceptron (MLP) neural network with Wavelet Transform for feature extraction. Accurate and timely stroke detection is critical for effective treatment and improved patient outcomes in neuroimaging. While Convolutional Neural Networks (CNNs) are widely used for medical image analysis, their computational complexity often hinders deployment in resource-constrained clinical settings. In contrast, our approach combines Wavelet Transform with a compact MLP to achieve efficient and accurate stroke classification. Using the "Brain Stroke MRI Images" dataset, our method yields classification accuracies of 82.0% with the "db4" wavelet (level 3 decomposition) and 86.00% with the "Haar" wavelet (level 2 decomposition). This analysis highlights a balance between diagnostic accuracy and computational efficiency, offering a practical solution for automated stroke diagnosis. Future research will focus on enhancing model robustness and integrating additional MRI modalities for comprehensive stroke assessment.

## I. Introduction

Magnetic Resonance Imaging (MRI) is a modern and non-invasive imaging technique that has revolutionized medical diagnostics. By using strong magnetic fields and radio waves, MRI provides high-resolution images of the internal structures of the body without exposing patients to ionizing radiation [1]. This powerful tool has opened up new opportunities for early disease detection, precise anatomical visualization, and improved treatment planning across various fields of medicine, particularly in neurology.

There are several different types of MRI sequences, each optimized to highlight specific tissue characteristics. Among the most commonly used are T1-weighted, T2-weighted, and FLAIR sequences. In this study, we focus on Diffusion-Weighted Imaging (DWI) [2], a specific type of MRI that is particularly effective for detecting acute ischemic strokes. DWI allows for the identification of areas with restricted water diffusion, which is a key indicator of early-stage ischemic injury in brain tissue. As such, it has become an essential tool in the timely diagnosis and management of stroke patients.

In recent years, neural networks and other machine learning techniques have gained significant attention in the medical field due to their ability to analyze complex data and recognize patterns [3] that may not be easily detectable by humans. These models have been successfully applied to a variety of tasks such as disease classification, medical image segmentation, outcome prediction, and treatment planning. In the context of stroke and neuroimaging, neural networks can assist in the automatic detection of lesions, improve diagnostic accuracy, and accelerate clinical decision-making processes. Their ability to learn from large datasets makes them powerful tools for supporting physicians and enhancing the efficiency and precision of modern healthcare systems.

Many studies in the field of medical imaging have primarily focused on utilizing Computed Tomography (CT) for stroke detection and other neurological disorders due to its widespread availability, fast processing times, and effectiveness in detecting hemorrhagic strokes. However, while CT plays a crucial role in the initial diagnosis of strokes, it has limitations in detecting acute ischemic strokes, particularly in the early stages when the brain tissue changes are subtle. As a result, researchers and clinicians are increasingly recognizing the advantages of Magnetic Resonance Imaging (MRI) in providing superior soft tissue contrast and its ability to detect ischemic changes much earlier. Despite the promise of MRI, it has not been as widely explored for automated stroke detection in comparison to CT. There is a growing need to shift the focus towards more advanced MRI techniques, like Diffusion-

Weighted Imaging (DWI), for better stroke diagnosis and management.

In terms of medical image processing, Convolutional Neural Networks (CNNs) have been the dominant tool in many studies [4], particularly in automating the analysis of MRI and CT images. CNNs excel in extracting spatial features and have demonstrated significant success in various imaging tasks such as image classification and segmentation. However, in our approach, we introduced a combination of neural networks with Wavelet Transform for feature extraction. The Wavelet Transform is particularly effective for capturing both frequency and spatial domain information, allowing for more detailed analysis of the image's multi-resolution features. This combination enables our model to better handle the complexity of MRI images, especially in detecting subtle ischemic changes that may be missed by CNNs alone. By integrating wavelet transform, we aim to enhance the model's sensitivity and accuracy in stroke detection, providing a more robust solution for clinical applications.

This paper presents a lightweight framework for classifying brain stroke types from Diffusion-Weighted Imaging (DWI) MRI scans using a Multi-Layer Perceptron (MLP) neural network with Wavelet Transform for feature extraction. It compares the performance of different wavelet configurations ("db4" and "haar") in terms of classification accuracy and computational efficiency, providing insights into their suitability for clinical applications. This framework demonstrates the potential of integrating wavelet-based feature extraction with compact neural networks for automated stroke diagnosis in resource-constrained settings.

The remainder of the paper is organized as follows: Section II reviews related work on neural networks and feature extraction techniques in medical imaging. Section III provides background information on Wavelet Transform and MLP neural networks to establish the foundational understanding necessary for addressing the problem. Section IV details the proposed framework and experimental setup, while Section V concludes the paper.

## II. RELATED WORKS

The integration of machine learning techniques, particularly neural networks, in medical imaging has significantly advanced the classification and diagnosis of neurological disorders. Numerous studies have explored the application of wavelet transforms combined with neural networks to enhance the accuracy of disease detection in MRI brain images.

In the context of Alzheimer's disease detection, researchers have employed wavelet transformation energy features extracted from structural MRI images [5], utilizing nearest neighbor classifiers to distinguish between Alzheimer's patients and healthy controls [6]. Similarly, a dual-tree complex wavelet transform combined with principal component analysis and feed-forward neural networks has been proposed, achieving notable accuracy in differentiating Alzheimer's disease from healthy states [7].

For epilepsy detection, studies have introduced wavelet-based deep learning approaches that eliminate the need for manual feature extraction [8]. These methods have demonstrated improved accuracy on smaller datasets compared to traditional machine learning algorithms [9]. Additionally, attention-based wavelet convolutional neural networks have been developed for epilepsy EEG classification, achieving state-of-the-art results [10].

Brain tumor classification has also benefited from the combination of wavelet transforms and neural networks. Researchers have utilized wavelet texture features and optimized artificial neural networks to recognize tumors from multimodal MRI images [8,11]. Furthermore, deep wavelet auto-encoder models have been proposed for brain tumor detection and classification, demonstrating high accuracy and low validation loss [12].

In terms of stroke classification using neural networks, several studies have shown promising results. A study using a modified MobileNetV2 architecture achieved 96% accuracy (κ: 0.92) for ischemic stroke detection on DWI images and 93% (κ: 0.895) for vascular territory classification [13]. Another investigation employing an Enhanced Convolutional Neural Network (CNN) reported 98.4% accuracy in distinguishing between ischemic and hemorrhagic strokes from MRI scans, outperforming traditional classifiers like SVM and standard ANN [14]. Hybrid models such as OzNet, which combines CNNs with machine learning algorithms, achieved 87.19% accuracy (AUC: 0.9488), with further improvements (up to 97.24%) when combined with Support Vector Machines [15, 16].

While these studies underscore the efficacy of neural networks in stroke classification, and of combining wavelet transforms with neural networks for various neurological disease classifications, there is a notable gap in applying this combination specifically to stroke detection. In particular, the use of different wavelet transforms in conjunction with neural networks for distinguishing between multiple stroke types—such as ischemic, hemorrhagic, and normal—using DWI MRI images has not been extensively explored.

Our research aims to address this gap by implementing wavelet transform techniques (Haar and Daubechies-4) alongside multilayer perceptron (MLP) neural networks to classify normal, ischemic, and hemorrhagic stroke cases using DWI MRI scans. This approach seeks to enhance classification accuracy and provide a robust framework for practical stroke diagnosis.

## III. BACKGROUND

### A. Neural Networks

Neural networks have become a cornerstone in medical image analysis due to their capacity to approximate complex, high-dimensional mappings between inputs and outputs [3]. These models are composed of layers of interconnected nodes, where each layer applies a non-linear transformation to its inputs [17], enabling the extraction of hierarchical representations from raw data. In contrast to traditional algorithms that rely on hand-engineered features, neural networks—particularly feedforward architectures like multilayer perceptron (MLP) and deep convolutional neural networks (CNNs)—learn feature representations directly from the data, optimizing performance for classification, segmentation, and detection tasks.

In medical imaging, neural networks offer several advantages, including robustness to noise, adaptability to heterogeneous data, and scalability with increasing dataset size. Their ability to uncover subtle patterns and anomalies in complex datasets, such as MRI or CT images, has led to significant improvements in diagnostic accuracy across various conditions, including brain tumors, Parkinson's disease, and cerebrovascular disorders. While CNNs dominate many image analysis tasks due to their spatial feature learning capabilities, other architectures like multilayer perceptron (MLP) are also widely used, particularly when working with pre-extracted features or lower-dimensional data representations.

### B. Wavelet transform

Wavelet transform is a mathematical technique used to analyze signals or images at different scales and resolutions [18]. Unlike traditional Fourier transforms, which decompose signals into sine and cosine functions representing only frequency components, wavelet transforms provide both frequency and spatial information. This is achieved through the use of wavelets, which are small oscillating functions that are localized in both time and frequency. The mathematical operation of wavelet transform involves applying these wavelets to the signal or image to break it down into several frequency bands at various resolutions.

Mathematically, the continuous wavelet transform (CWT) of a signal f(t) is given by:

$$W(a,b) = \int_{-\infty}^{\infty} f(t)\psi^*\left(\frac{t-b}{a}\right)dt$$

where $\psi$ is the wavelet function, a is the scale factor, and b is the translation parameter. The scale factor a controls the resolution of the wavelet, with larger values corresponding to coarse-scale features and smaller values corresponding to fine-scale features. The symbol $\psi^*$ represents the complex conjugate of the wavelet function, which is used in the transform to ensure proper mathematical handling of energy and phase—especially when complex-valued wavelets are involved. By applying this transformation, the image is decomposed into high- and low-frequency components, providing a multi-scale representation of the data.

In medical imaging, wavelet transform is commonly used for feature extraction, particularly in detecting subtle variations in tissue structures. For example, in MRI scans, wavelet transform can highlight fine details such as small lesions or ischemic areas in brain tissue, which are crucial for early diagnosis. This multi-resolution approach ensures that both coarse structures and fine details are captured, making it valuable for detecting conditions such as stroke, Parkinson's disease, and brain tumors.

One of the key advantages of wavelet transform is its ability to reduce noise and artifacts while preserving important image features. The process of wavelet decomposition allows for noise suppression while maintaining critical diagnostic information. This is particularly important in medical imaging, where noise can obscure vital details and lead to misdiagnosis. Moreover, wavelet transform helps improve the computational efficiency of subsequent analysis by reducing the dimensionality of the data while retaining essential features. This makes it easier for neural networks to process and interpret the data, ensuring that important diagnostic information is not lost.

## IV. OUR APPROACH

In this study, we developed a lightweight and efficient method for classifying brain stroke types from Diffusion-Weighted Imaging (DWI) MRI images using a neural network. We utilized the "Brain Stroke MRI Images" dataset from Kaggle, which includes MRI scans from patients with normal, ischemic, and hemorrhagic stroke conditions, focusing exclusively on DWI images due to their effectiveness in early ischemic stroke detection.

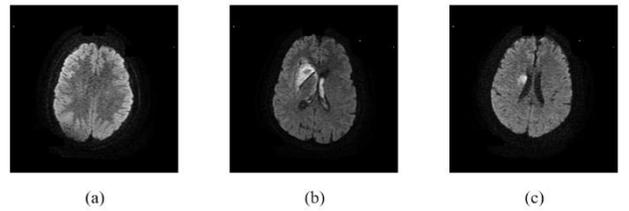

*Figure 1: Brain MRI scans: (a) Normal brain, (b) Hemorrhagic stroke showing blood accumulation, and (c) Ischemic stroke with visible tissue damage due to reduced blood flow.*

The raw DWI images required preprocessing to remove diagnostic text labels and annotations. We applied a masking technique in Python to clean the images and resized them to 256×256 pixels for consistency [19]. To enhance model robustness, we augmented the dataset with random rotations (up to 10 degrees), a 50% chance of horizontal flipping, and random brightness adjustments. Direct use of raw pixel values (over 65,000 features) was computationally inefficient and ineffective for capturing diagnostic patterns. Instead, we employed Wavelet Transform to extract a compact 128-dimensional feature vector per image, capturing both spatial and frequency information critical for identifying stroke-related patterns.

We selected Wavelet Transform over Fourier Transform for its ability to provide localized spatial-frequency analysis, essential for detecting subtle ischemic and hemorrhagic changes. Two wavelet configurations were tested: the "haar" wavelet with level 2 decomposition and the "db4" wavelet with level 3 decomposition. These configurations were chosen to balance feature richness with computational efficiency, enabling deployment in resource-constrained clinical settings.

The extracted 128-dimensional feature vectors were fed into a Multi-Layer Perceptron (MLP) neural network for classification. The MLP architecture comprises an input layer with 128 neurons, followed by two hidden layers with 128 and 64 neurons, respectively, both utilizing ReLU activation and initialized with a normal kernel initializer. Batch normalization and dropout layers (30% and 10% rates, respectively) are incorporated after each hidden layer to enhance training stability and prevent overfitting. The architecture concludes with an output layer of 3 neurons employing Softmax activation, corresponding to the diagnostic classes of normal, ischemic, and hemorrhagic stroke. This compact yet effective architecture mapped wavelet-extracted features to diagnostic outcomes efficiently, avoiding the computational complexity of deeper models.

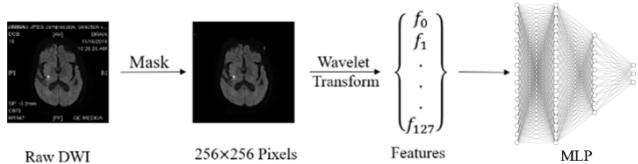

Figure 2: The procedure of decision making: Raw DWI data is masked, transformed using Wavelet Transform into 256*256-pixel features, and then fed into a Multi-Layer Perceptron (MLP) for analysis.

We evaluated the performance of the two wavelet configurations, "haar" (level 2) and "db4" (level 3), using precision, recall, F1-score, and accuracy metrics for the classification of hemorrhagic stroke, ischemic stroke, and normal cases. Additionally, we calculated the overall sensitivity for detecting stroke cases (both hemorrhagic and ischemic) to assess the model's ability to identify strokes, even if the specific stroke type was misclassified. This is critical in clinical settings, as missing a stroke diagnosis (false negative) is more detrimental than misclassifying the stroke type.

Table 1: Performance Metrics for db4 Wavelet (Level 3)

| Diagnosis Type | Precision (%) | Recall (%) | F1-score |
|---|---|---|---|
| Hemorrhagic Stroke | 90.00 | 53.00 | 67.00 |
| Ischemic Stroke | 80.00 | 48.00 | 60.00 |
| Normal | 81.00 | 97.00 | 88.00 |

For the db4 wavelet (level 3), the model achieved a train accuracy of 77.40%, a validation accuracy of 81.06%, and a test accuracy of 82.00%. The overall stroke sensitivity, representing the percentage of stroke cases correctly identified as stroke regardless of type, was 51.16%.

Table 2: Performance Metrics for Haar Wavelet (Level 2)

| Diagnosis Type | Precision (%) | Recall (%) | F1-score |
|---|---|---|---|
| Hemorrhagic Stroke | 88.00 | 72.00 | 79.00 |
| Ischemic Stroke | 74.00 | 61.00 | 67.00 |
| Normal | 87.00 | 94.00 | 90.00 |

For the Haar wavelet (level 2), the model achieved a train accuracy of 83.20%, a validation accuracy of 83.71%, and a test accuracy of 86.00%. The overall stroke sensitivity, representing the percentage of stroke cases correctly identified as stroke regardless of type, was 67.44%.

The Haar wavelet (level 2) outperformed the db4 wavelet (level 3) with a test accuracy of 86.00% compared to 82.00% and an overall stroke sensitivity of 67.44% compared to 51.16%. The Haar wavelet demonstrated higher recall and F1-scores for hemorrhagic and ischemic stroke classifications, indicating better performance in identifying stroke cases. Notably, the Haar wavelet achieved a significantly higher recall for hemorrhagic stroke (72.00% vs. 53.00%) and ischemic stroke (61.00% vs. 48.00%), which contributed to its superior overall stroke sensitivity. The db4 wavelet, however, showed slightly better precision for hemorrhagic stroke (90.00% vs. 88.00%) and ischemic stroke (80.00% vs. 74.00%) and a higher recall for normal cases (97.00% vs. 94.00%). However, since recall for stroke cases — and more broadly, the overall sensitivity for detecting any type of stroke — is of paramount importance in medical diagnostics, the Haar wavelet, which also achieved a higher test accuracy, is a more suitable choice. The Haar wavelet's simpler structure likely contributed to its robustness and better generalization, while the db4 wavelet's higher decomposition level may have introduced noise or redundant features, reducing its effectiveness. Given its higher accuracy and sensitivity, the Haar wavelet is more suitable for resource-constrained clinical applications where reliable stroke detection is critical.

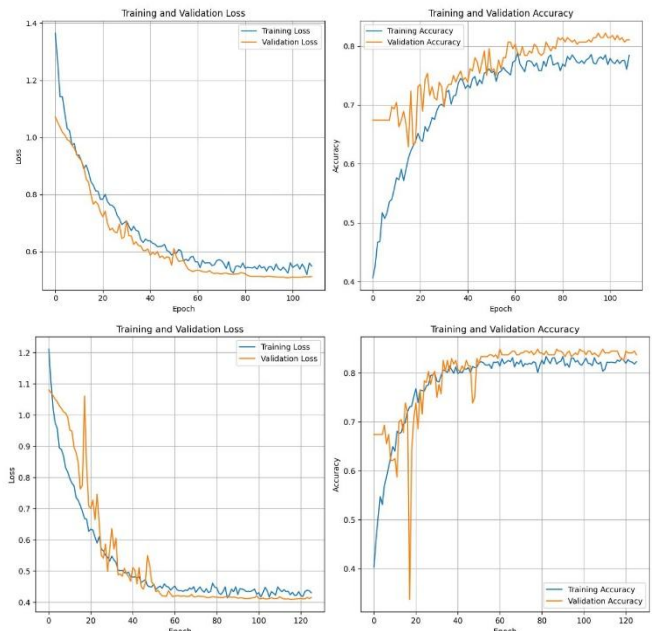

*Figure 3: Loss and accuracy curves over epochs for two models. (Top) Loss and accuracy plot of the db4 model with an accuracy of 82.00%. (Bottom) Loss and accuracy plot of the Haar model with an accuracy of 86.00%.*

## V. CONCLUSION

This study presents a lightweight and efficient method for classifying brain stroke types from Diffusion-Weighted Imaging (DWI) MRI scans using a combination of Wavelet Transform and Multi-Layer Perceptron (MLP) neural networks. Our approach demonstrated that wavelet-based feature extraction can significantly reduce the dimensionality of MRI data while preserving critical diagnostic information, allowing for accurate classification without the computational overhead of deep convolutional networks.

By experimenting with two wavelet configurations—Daubechies 4 (db4) and Haar—we showed that different wavelet types can influence classification performance, with the Haar wavelet at level 2 decomposition achieving the best accuracy of 86.00%. This highlights the potential of tailoring wavelet parameters to optimize diagnostic outcomes in medical imaging tasks.

Compared to more complex neural architectures such as CNNs and hybrid deep learning models previously used in stroke classification, our method offers a balanced solution that prioritizes both accuracy and resource efficiency. While other models have reported higher accuracies, such as MobileNetV2-based architectures achieving up to 96%, they often require significantly more computational power and are less suitable for embedded clinical applications.

Our results suggest that integrating wavelet transforms with compact neural network architectures can serve as a practical and scalable approach for automated stroke diagnosis, especially in settings where computational resources are limited. Future work will explore the incorporation of additional imaging modalities, larger and more diverse datasets, and the potential of hybrid models combining the interpretability of shallow networks with the feature extraction power of deeper ones [20] to further enhance diagnostic precision and robustness